\documentclass{blois}

\def\be{\begin{equation}}
\def\ee{\end{equation}}
\def\bea{\begin{eqnarray}}
\def\eea{\end{eqnarray}}

\usepackage{soul}  

\newcommand{\Photo}{\includegraphics[height=35mm]{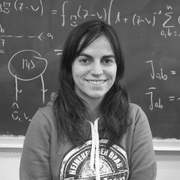}}

\begin{document}
\vspace*{4cm}
\title{Study of VBF/VBS in the LHC at 13 TeV,  the EFT Approach.}

\author{ Raquel Gomez Ambrosio }

\address{Dipartimento di Fisica Teorica, Universit\`a di Torino and INFN Sezione di Torino \\ 
Via Pietro Giuria 1, 10125 Turin, Italy }

\maketitle\abstracts{
We review the current status of vector boson fusion and vector boson scattering studies in the frame of Run-2 of LHC. These processes are equally interesting and challenging, and new ideas have to be applied for their experimental analysis. These channels are optimal for the search of new physics, and therefore we discuss two promising techniques for this purpose: standard model effective field theory and pseudo observables. Regarding their implementation in the analyses, we also discuss briefly how these techniques should be included in the Monte Carlo generators. 
}

\section{Introduction: Why Study VBF/VBS at LHC}

Vector Boson Fusion (VBF) is the subleading channel for Higgs production at LHC.  More concretely we call VBF to the production of a Higgs boson, in association with two hard jets, in the forward and backward regions of the detector. However, this process can only be understood in the frame of the vector boson scattering (VBS), which involves other Higgs production channels (VH associated production) and processes with no Higgs bosons (VV $\rightarrow$ VV).  


Historically, the family of VBS processes has illustrated a paradigmatic problem in theoretical particle physics: violation of unitarity. If there is something that we know for sure in quantum field theory is that unitarity,  representing probability, has to be conserved.  Without the Higgs boson, the cross section for vector boson scattering would grow proportionally with the center of mass energy 

\vspace{-0.5cm}
\begin{equation}
\sigma_{V_L V_L \rightarrow V_L V_L}  \propto   \left[ -s -t + \underbrace{\frac{s^2}{s - m_H^2} + \frac{t^2}{t - m_H^2}}_{\rm{Higgs \, \,  contribution}} \right]
\end{equation}

There is a subtlety here to be noted: although the Higgs ensures the convergence of the cross section when $s \rightarrow \infty$, we still have very little information about the behaviour in the intermediate energy regime, where phenomena like delayed unitarity\cite{Ahn:1988fx}  could possibly hide the portals to new physics.

From the experimental perspective,  all VBS processes have a very clean signature. There are two very energetic forward jets, which can be tagged,  and  the products of the decay  lay in the central region of the detector, well separated from the jets. However a challenge still exists, on separating the Higgs signal from the background, since they both share these particular characteristics. This is one of the tasks to be addressed on the LHC run-2 dataset, possibly using a MELA\footnote{Matrix Element Likelihood Approach} approach.  A further difficulty for the experimental analysis  is the fact that the final decays can be of the type $ZZ \rightarrow 4 \ell$, $ ZZ \rightarrow 2 \ell 2 \nu$ or $W^{\pm} W^{\pm} \rightarrow \ell \nu \ell \nu$. These processes are quite similar from the theory point of view and in fact interdependent, since all are needed to satisfy gauge invariance, but completely different for the experimental reconstruction and analysis.



\section{The Importance of NLO EW corrections}

When we talk about NLO corrections in LHC we are often only including QCD ones. However there is evidence that electroweak corrections should not be neglected at the current precision. The VBF total cross section is: 

\begin{equation}
\sigma_{\rm{VBF}} = \sigma^{\rm{DIS}}_{\rm{NNLOQCD}} (1 + \delta_{EW}) + \sigma_{\gamma} = \underbrace{1241.4 \, \, \rm{fb^{-1}}}_{7 \, \, \rm{TeV}} / \underbrace{4277.7 \, \, \rm{fb^{-1}}}_{14 \, \, \rm{TeV}} \, \pm \Delta
\end{equation}

were $\delta_{\rm{EW}}$ has been calculated\cite{Denner:2016kdg} to be $-4.4\%$(7 TeV) and $-5.4\%$(14 TeV), and reaching $-6.9\%$ if we restrict to the fiducial cross section.

\subsection{\texttt{Collier} and \texttt{Recola}  }

At the computational level, two very promising pieces of software were released recently, namely RECOLA\cite{Actis:2016mpe} which is able to produce matrix elements for NLO EW corections, and COLLIER\cite{Denner:2016kdg} which can perform the phase space integration of such elements. These two projects open the door to the implementation of NLO EW corrections in the event generators used for the experimental analyses\footnote{Note that such corrections are already included in the Monte Carlo generators that theorists use privately.}.

\section{Effective Field Theory}

In order to look for small deviations from the SM\footnote{In the energy regime accessible to us, there are enough hints to assume such deviations will be small} but keeping the quantum field theory behind consistent (with the assumption that QFT must be the underlying theory governing new physics), the most reasonable path to follow is that of effective field theory (EFT).

EFT comes in two possible flavors, the top-down approach wich is strictly model dependent. And the relatively model independent bottom-up approach. In the top-down approach we start by an ultraviolet (UV) theory, find its low-energy behaviour, and try to match it to the SM in that regime. This is a quite straightforward technique, and it can provide with tests and predictions for concrete BSM models. First applying the path integral formalism and the covariant derivative expansion\cite{Henning:2014wua} and then using the renormalization group equations, one can match the Wilson coefficients of the UV theory with those of the SM.  This procedure has to be applied carefully though, since one might be neglecting mixing terms\cite{delAguila:2016zcb,Boggia:2016asg} between light and heavy particles if we try to apply the procedure in a generalized way.  Alternatively, the bottom-up approach starts from the SM and tries to build a bridge \emph{towards} the UV regime. This is, it does not provide with concrete predictions for BSM scenarios but rather with a framework for the study of new physics in the regime accesible from the SM.  We will address this approach in detail in the following.

An EFT for the standard model can be parametrized as: 

\begin{equation}
\mathcal{L} = \mathcal{L}_{SM} + \sum_{d > 4} \sum_{n} \frac{c_n}{\Lambda^{2d}} \mathcal{O}_n^{(d)}  =  \mathcal{L}_{SM} + \sum_{n} \frac{c_n}{\Lambda^2} \mathcal{O}_n^{(6)} + \sum_{m} \frac{c_m}{\Lambda^4} \mathcal{O}_m^{(8)} + \dots 
\end{equation}

where  $\mathcal{O}_n^{(d)}$ are operators of dimension $d>4$, representing new interactions but built completely from standard model fields\footnote{Think for instance on Fermi's operator, representing a direct interacion between four fermions without Gauge bosons mediating.}    and $\Lambda$ is  \emph{a scale of new physics} up to which the EFT is valid. Observe that $\Lambda$ is not necessarily the Planck scale, it might be that on top of the EFT we need another EFT, or a series of them, to extend its regime of validity. 

This Lagrangian is renormalizable, order by order in $\Lambda$ and in fact it has to be renormalized in order to make accurate predictions, since the self energies, tadpoles and counterterms will be modified\cite{Ghezzi:2015vva} with respect to the standard model ones. In particular, for the search for deviations of the SM in the energy regime accesible from LHC, it is enough to implement and renormalize dimension 6 operators. The full set of gauge-invariant, independent, dimension-6 operators can be parametrized in the so called \emph{Warsaw basis}\cite{Grzadkowski:2010es}. Analogously to the Lagrangian, the amplitude squared can be expanded as: 

\begin{equation}
\vert \mathcal{A} \vert^2 = \Bigg\vert \mathcal{A}_{SM} + \frac{\mathcal{A}^{(6)}}{\Lambda^2}  +  \frac{\mathcal{A}^{(8)}}{\Lambda^4} \Bigg\vert^2  =  \mathcal{A}_{SM}^2 + \frac{2}{\Lambda^2} \Re( \mathcal{A}_{SM} \mathcal{A}^{(6)}) + \mathcal{O}(\Lambda^{-4})
\end{equation}

In terms of Feynman diagrams this means that, to be consistent, we should only consider those diagrams with only Standard Model vertices ($\mathcal{A}_{SM}$) and those diagrams with one dimension 6 insertion ($\mathcal{A}_{SM} \mathcal{A}^{(6)}$). For terms with two dimension 6 insertions to be considered, we should as well include those with one dim 8 insertion (i.e. order $\Lambda^{-4}$ in $\vert \mathcal{A} \vert^2$). Nevertheless the former should be calculated as an estimate for the theoretical uncertainty. 

%

\subsection{Predictions within EFT}

After the framework has been designed, we have to figure out which are the most interesting or urgent applications. The goal of LHC in the long term is to find new physics and answer the open questions in high energy physics, however in the short term, the paths towards these goals are several. 

The EFT at dimension 6 naturally adds 59 new operators to the SM Lagrangian, and therefore, 59 new Wilson coefficients that have to be fitted. The main open question right now is if it makes sense to try and fit these coefficients independently, this is, varying them one by one and seeing if they hint at some deviation from the SM, or alternatively, if we should try to do a global fit. The latter option is more reasonable since the coefficients are correlated in non obvious ways, but it is also harder to implement. We can go one step further and discuss if it makes sense to look at Wilson coefficients \emph{at all}, since they are not observable, nor independent and, when compared to other quantities like scattering amplitudes, they seem less natural.

\subsection{A complement to EFT: Pseudo Observables}


A natural bridge between EFT and experimental data is the pseudo observable (PO) approach. POs are strictly related to EFT parameters, but they are not the same thing. They are universal quantities, defined from a theoretical perspective, but with a well defined experimental meaning. We can divide them into two big groups: effective on-shell couplings (equivalent to the $\kappa$ framework, and restricted to lowest order) and \emph{ideal observables}\cite{Greljo:2015sla}, such as partial decay widths, electroweak precision data, and differential cross-sections. 

POs are extracted from a decomposition of the scattering amplitude, followed by a momentum expansion (therefore assuming there are no new light particles). Using analiticity, unitarity and crossing symmetry arguments, one can define these quantities to be the coefficients of such a decomposition. 
The crossing symmetry argument is particularly interesting, since it relates the POs from the production and decay modes, lowering the theoretical uncertainty (and amount of work) when we carry the analysis for a complete $pp \rightarrow H \rightarrow X$ process.

\section{Implementing NLO EFT corrections in the Monte Carlo generators}

As we addressed in the introduction, the set of VBS Feynman diagrams only respect gauge invariance if they are considered together. And this as well is the main feature we have to take into account when implementing NLO EW corrections and NLO EFT in the Monte Carlo generators. 

The algorithms for the inlcusion of QCD NLO corrections in these generators are such that each Feynman diagram is computed and integrated over the phase space independently, and the different contributions are summed up in the end. However this method is not the best one to use in the case of electroweak processes.

Firstly, electroweak vertices (and EFT-EW ones) are of a much more complicated nature. Due to the presence of masses in the propagators, the loop integrals are more time- and resource-consuming, and these are the main variables that one needs to optimize when designing a MC generator. Secondly, and more important, there is the question of being rigorous: since we know from the SM that gauge cancellations have to happen, it does not look right to calculate the contributions of different non-gauge invariant diagrams separately using numerical techniques, and then sum them up hoping that those cancellations calculated numerically will cancel, specially given the fact that these spurious contributions are relatively large in the EW case

\section{Acknowledgements}
Work supported by the Research Executive Agency (REA) of the European Union
under the Grant Agreement PITN-GA-2012-316704 (HiggsTools).

\section*{References}

\bibliography{mybib}{}
\end{document}